\documentclass[letterpaper,twocolumn,10pt]{article}
\usepackage{usenix,epsfig,endnotes}
\usepackage{caption}
\usepackage{subcaption}
\begin{document}

\newcommand{\repeatthanks}{\textsuperscript{\thefootnote}}

\date{}

\title{\Large \bf Towards Perceived Security, Perceived Privacy, and the Universal Design of E-Payment Applications}


\author{
{\rm Urvashi Kishnani, Isabella Cardenas\thanks{These authors contributed equally to this project}, Jailene Castillo\protect\footnotemark[1], Rosalyn Conry\protect\footnotemark[1],}\\
{\rm Lukas Rodwin\protect\footnotemark[1], Rika Ruiz\protect\footnotemark[1], Matthew Walther\protect\footnotemark[1], and Sanchari Das} \\
University of Denver
}


\maketitle

\pagestyle{empty}

\subsection*{Abstract}
With the growth of digital monetary transactions and cashless payments, encouraged by the COVID-19 pandemic, use of e-payment applications is on the rise. It is thus imperative to understand and evaluate the current posture of e-payment applications from three major user-facing angles: security, privacy, and usability. To this, we created a high-fidelity prototype of an e-payment application that encompassed features that we wanted to test with users. We then conducted a pilot study where we recruited $12$ participants who tested our prototype. We find that both security and privacy are important for users of e-payment applications. Additionally, some participants perceive the strength of security and privacy based on the usability of the application. We provide recommendations such as universal design of e-payment applications.

\section{Introduction}

E-payment applications (such as Google Pay, Apple Pay, Cash App, PayPal, or Venmo) allow for contactless transactions, giving people opportunities to buy goods or services, or transfer money when traditional payment methods are unsafe or unavailable. According to a $2019$ report by Mckinsey and Company, about $77\%$ of people made a mobile payment in the year $2019$ ~\cite{mckinsey_company_2019}. As a major financial system in modern economy, e-payment systems must aim to be absolutely secure with financial information and accounts. Users of e-payment systems not only face certain financial risks, but are also prone to privacy risks. This is because most e-payment applications require critical information such as Personally Identifiable Information (PII) of the users, including name, date of birth, address, and banking information. Thus, users are susceptible to potential loss of PII, introducing the threat of identity theft~\cite{ho_price_2020}. Additionally, usability and design of e-payment applications is also equally important, as prior studies have shown that usability of mobile application plays a role in the perceived security and privacy understanding by the user~\cite{ozkan_facilitating_2010}.

\section{Methodology}
\label{methodology}

\subsection{Prototype Building}
We make one comprehensive prototype to test various functionalities and behaviors that were previously studied separately (such as~\cite{de_assessment_2015, kim_empirical_2010, ho_price_2020, mingyen_teoh_factors_2013, ogbanufe_comparing_2018}). Unlike most payment apps, our prototype presents the option to view the privacy policy prominently. Moreover, instead of a lengthy policy that users usually do not read, we present a short simplified privacy policy screen to gauge the engagement of users. Lastly, unlike most other payment apps, we provide mandatory Multi-Factor Authentication (MFA) options to our users at the time of login and registration. Additionally, our prototype incorporates four different MFA options and gives users the choice to select their preferred method, thereby improving security while still allowing users to feel in control. The prototype is built using Figma, which is free for students and educators~\footnote{https://www.figma.com/education/}. The prototype has several interlinked pages implementing security, privacy, and usability concepts. The prototype consists of a total of $17$ screens. The prototype is built to only be intractable with buttons and does not allow for any text-based input. This ensures that we do not accidentally collect any personal data from the participants during our study.

\begin{figure}
     \centering
     \begin{subfigure}[b]{0.1\textwidth}
         \centering
         \includegraphics[width=\textwidth]{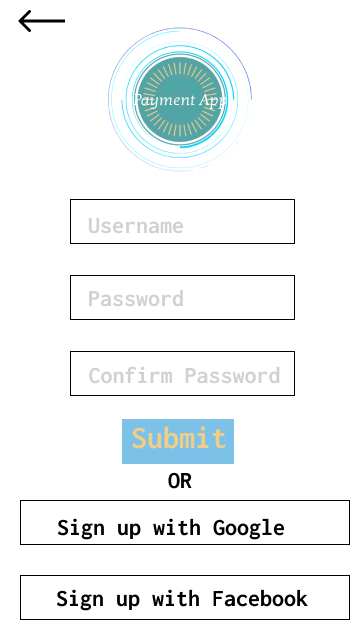}
         \caption{}
         \label{fig:registration}
     \end{subfigure}
     \hfill
     \begin{subfigure}[b]{0.1\textwidth}
         \centering
         \includegraphics[width=\textwidth]{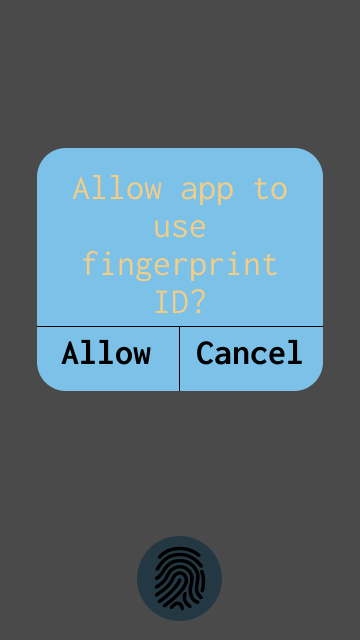}
         \caption{}
         \label{fig:fingerprint}
     \end{subfigure}
     \hfill
     \begin{subfigure}[b]{0.1\textwidth}
         \centering
         \includegraphics[width=\textwidth]{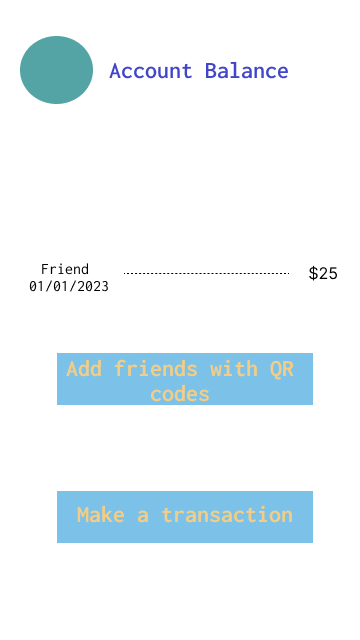}
         \caption{}
         \label{fig:homepage}
     \end{subfigure}
     \hfill
     \begin{subfigure}[b]{0.1\textwidth}
         \centering
         \includegraphics[width=\textwidth]{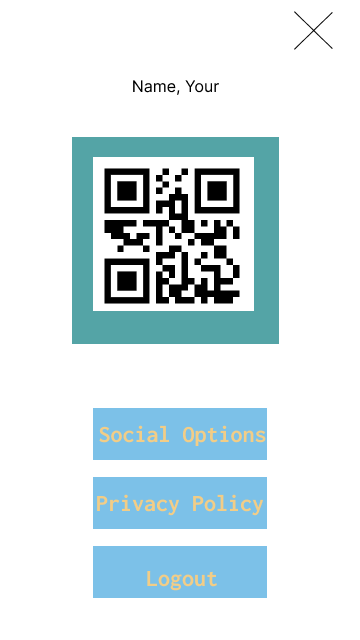}
         \caption{}
         \label{fig:settings}
     \end{subfigure}
    \caption{Screenshots of Prototype (Left-to-Right: Registration Page, Permissions Page, Home Screen, Settings Page.)}
    \label{fig:screenshots}
\end{figure}

\subsection{User Study}

\textbf{Recruitment and Eligibility:}
Recruitment information including an eligibility criteria and link to a pre-screening survey was distributed using flyers, emails, social media posts. To be eligible to participate in the study, participants had to meet the following requirements: Participant must (1) be able to come to the university campus to perform the in-lab study, (2) have experience in using at least one e-payment or banking application, and (3) be $18$ years of age or older and must reside in the US. Those participants that consented to the study and met all the requirements proceeded to fill in our pre-screening survey based on survey instrument proposed by Rajivan et al.~\cite{rajivan2017factors}. From the $16$ respondents, we selected $N=12$ participants who were then invited to an in-person study where they first tested and interacted with a prototype and then answered questions to a semi-structured interview. The study with each participant took $20-30$ minutes. 

\textbf{Interview and Qualitative Analysis:}
The study was conducted in the university campus by the primary researcher of this work. At the time of the interview, the participant were given an overall briefing about the task, and then tested and interacted with the application prototype. After this, they were asked to answer questions to a semi-structured interview about the usability, privacy, and security features of the application. There was no compensation for participation. To facilitate the qualitative analysis of the interviews, each interview was audio recorded if appropriate consent was given by the participant. These interviews were then transcribed using Zoom and Google Pixel $7$ Pro's automatic live transcription services to convert the audio to textual context. Two researchers open coded two interviews to generate keywords and phrases of interest and grouped these codes on two-levels to form the codebook. Thereafter, we conducted a thematic analysis of the remaining transcribed interviews. 

\section{Results}
\label{result}

Table~\ref{table:demographics} summarizes the demographics of our participants. Our participants, being university students, represented a relatively younger population. The distribution of education level is in agreement with the participants' ages. We discuss themes and corresponding sub-themes from our thematic analysis below.

\begin{table}[]
\caption{Participant demographics (NB: Non-Binary, NS: Not Specified)}
\label{table:demographics}
\begin{tabular}{lllll}
\hline
\textbf{ID} & \textbf{Age} & \textbf{Gender} & \textbf{Education Level} & \textbf{Ethnicity}     \\
\hline
P01             & 18-24  & Male            & High School              & White                  \\
P02             & 18-24  & Male            & High School              & White                  \\
P03             & 18-24  & Male            & Diploma                  & White                  \\
P04             & 18-24  & Male            & Bachelors          & White                  \\
P05             & 18-24  & Male            & Diploma                  & White                  \\
P06             & 18-24  & Female          & High School              & White                  \\
P07             & 18-24  & Female          & Diploma                  & White                  \\
P08             & 25-30  & Male            & Bachelors          & NS          \\
P09             & 18-24  & NB      & Bachelors          & White                  \\
P10             & 18-24  & Female          & Bachelors          & White                  \\
P11             & 18-24  & Female          & Bachelors          & Asian                  \\
P12             & 18-24  & Male            & High School              & Hispanic              \\
\hline

\end{tabular}
\end{table}

\subsection{Security}

\textbf{Perceived Security: }
Seven participants admitted that they find it difficult to comment on the security of the application and are unsure whether it is secure or not. Two participants mentioned that the blue colored theme represented security to them. Participant P02 mentioned that the more complicated an application, the more problems that can create. Participants also mentioned how the usability of the application impacted their view on the security.



\textbf{Multi-factor Authentication: }
Our prototype and interview questions focused on two applications of Multi-Factor Authentication (MFA). The first is during registration/login, and the second is before sending money. Eight participants felt that the availability of various MFA options during login meant that the application was more secure. Three participants mentioned that MFA security was more important than the convenience of a faster login. Five participants mentioned that they preferred OTP options, and one explicitly mentioned that they liked OTP over biometrics. Three participants said that they found the use of FaceID to be concerning for payment applications and two expressed that they had faced some usability barrier in using FaceID. On the other hand, three participants said they preferred biometric options over OTP, three participants said that they find FaceID convenient, and one participant explicitly said that they did not like OTP. Moreover, three participants said they find app-based MFA options (like Duo) annoying. 





\textbf{Login Notifications: }
Most participants responded positively to login notifications. Nine participants said they really liked them and found them to be important, two participants found them annoying and one participant said that they never received them. Regarding actions that the participants take when they receive login notifications, seven participants also mentioned that they don't take any action if they know the notification was triggered because of them or if they see the notification from the area that they are in. Seven participants also mentioned that if they recognize that the login was not made by them, they take action to update their information or change their passwords. One participant offered additional recommendation to enhance the functionality of login notifications.


\textbf{Adding Beneficiaries using QR Codes: }
Six participants appreciated the additional layer of identity verification that adding beneficiaries using QR codes provided. Five mentioned that they have used it a few times and two said that they use it frequently. However, four participants said that this was a new concept to them. But three mentioned that even though they haven't use this, they can understand how it can be beneficial. Participant in favor of QR Codes mentioned this help increase usability of the application.


\subsection{Privacy}

\textbf{Perceived Privacy: }
Five participants said that they found the app was built while keeping users privacy in mind. They mentioned different reasons for this opinion. Some said that they felt the option to add beneficiaries by QR code meant that the app was built with a privacy lenses. Some said that the prompts provided by the application for the prompts for permissions requested for camera of fingerprint indicated that the app respected user's privacy. One participant said that the app respected their privacy as they were not asked to write messages while sending money as they found this option intrusive in Venmo. Three participants said that they did not find this application was built for user's privacy and they mentioned various reasons for this. Some said that they found that there were no options to toggle privacy settings. Some participants mentioned that they required more transparency on the third parties that were involved. One participants said that they do not trust payment applications due to the financial interest of companies. Two participants said that there was no reason to not trust the app. We note that two participants found the app to be more private due to the authentication features, although that is more geared towards security. 



\textbf{Application Permissions: }
Six participants say that they scrutinize permissions overall. Participants mainly focused on three major permissions: location, contacts and camera. Two participants also mentioned that permission pop-ups are annoying but they find that they are okay for crucial applications like e-payment or banking. Four participants mentioned that the specifically scrutinize location permissions. One participant mentioned that they feel location is helpful to know when a transaction was made and can be used to prevent scams if we transaction was made in an area where the user was not present. Four participants mentioned that this scrutinized contacts. Two participant mentioned that they find sharing contacts is inconvenient especially because they have a large number of contacts and they don't want all of them to be imported into the payment application. Four, participants mentioned that this scrutinized camera and that they would allow camera for functional reasons if required by the application, like scanning QR code or taking pictures of cheque. Finally when asked about the permissions requested by the prototype, the participants felt that they were pretty standard and something that they would expect. 


\textbf{Privacy Policies: }
When asked about the participants experience with privacy policies in general, $9$ out of $12$, participants said that they do not read privacy policies. Only one participant mentioned that they do read privacy policy and that they understand that too. Three participants also mentioned that they do not understand privacy policies and three participants mentioned that they do understand privacy policies. When asked about the privacy policies of the prototype, eight participants mentioned that they just read, skimmed, or glanced over the policy for the prototype whereas five mentioned that they did not read the privacy policies for the prototype. When asked, what if they were easily able to find the privacy policy, six participants said that they were able to find that easily, whereas only one participant said that they felt the privacy policy option was hard to find or quite hidden.



\textbf{Social Features: }
Social features of payment applications, like Venmo which provides a social media-like feed of transactions, were mostly received with a negative sentiment. Six participants found this concept to be weird or strange and five participants said that they prefer private transactions. Four participants also mentioned they find social features to be intrusive and four participants said that there's no need to make payment apps like a social media app. Three participants said that it's strange to give access to contacts in the first place when it comes to payment applications. Two participants said that the default should be opt-out and the participants should have the option to opt-in if required, instead of it being the other way around. One participant mentioned that such social features can actually act as identity verification. Finally, two participants said that they had no opinions on social features for e-payment applications. 




\subsection{Usability}

\textbf{Overall User Experience: }
When it came to the overall user experience of using the prototype, nine participants said that they found it easy to use. Five said that the application was quite simple and five others said that they found it quite intuitive to use. One participant said that they neither liked nor disliked the overall usability of the application. Eight participants said that they like the usability factor of using QR codes as it make it much easier for them to find someone by scanning another person's QR code as compared to finding someone by name or typing complicated usernames. They also said that this would help avoid mistakes and would make sure that the money is sent to the right person. One participant pointed out that there was no option to withdraw money to their bank, and that this is one option that they use regularly and they said that it was missing from this prototype. 


\textbf{Navigation: }
When asked about the experience of navigating, the application nine participants found that it was simple and easy to navigate. They like having important features up front in the application, such as transactions on the home-screen immediately after logging in. However, some participants also felt that shortcuts like a pop-up menu under the profile picture, a navigation bar at the bottom with a home button would help make navigation easier. One participant mentioned that they were confused that after they registered, it went straight to the home screen instead of asking them to login again after registration. 


\textbf{Font: }
Only three participants said that the font on the prototype was easy-to-read. Two participant mentioned that they would change the font in a few places, for example, by adding outline to text to make it more legible. When asked about the option to change font size or type, most participants said that they wouldn't bother using such a feature. However, five participant did say that they would appreciate such a feature and two participants specifically mentioned that increasing the font size would be helpful to older adults or people with low vision, which would make the app more inclusive.



\textbf{Icons: }
Four participants said they enjoyed icons for navigability, but one warned that a lot of icons could make an app cluttered. With the consensus being that some amount of icons for important actions are best. For the prototype, four participants didn't notice many icons and said that more icons could be added. Two participants said that the icons could be made a little smaller. Participants also mentioned that the icons used should be universal and well known.

\section{Discussions and Implications}
\label{implication}

\textbf{Reduce Social Features:}
Most of our participants did not have positive opinions on Social features for E-Payment Applications where other user's can view their transactions. One participant also finds the concept of adding a note to each transaction to be intrusive. Users expressed disinterest in viewing transactions of others. One user suggested that such a concept can contribute to gamification of E-Payment applications where users may be pushed to send or receive more money. When asked about whether our participants used this option on an application like Venmo, all of our users said that they have not set their transactions to be public. Some users said that they only share their transactions with close family and friends. Social features like these also require users to share permission to their contacts, which allows apps to add ``friends" for the users. However, our participants also expressed concern with sharing all of their contacts with a payment application, and would prefer to only share selected contacts on need basis. Prior work has shown that social features of apps like Venmo can cause security issues~\cite{kraft2014security}. Our results of participants feeling indifferent or awkward about seeing other's transaction corroborates with prior research~\cite{caraway2017friends}.

\textbf{Fine-Grained Options for User’s Settings:}
Users like to be know that they are in control of their privacy, this makes them have more trust in the application that they are using. Most users expressed that multiple two-factor authentication options made them feel more secure. Some users preferred Biometric Authentication like Fingerprint and FaceID, over One Time Passwords (OTP) like Email or SMS, and others had the opposite preference. So having all these options presented to the user was received positively. Users also expressed concerns for a missing page in the prototype where they could fine-tune more settings. Further, one user said that giving permission to camera separately for each feature (like FaceID and scanning QR codes) was preferred, as compared to giving the permission one time for the entire app. Prior studies agree with user perception to be in favor of two-factor authentication as compared to single-factor authentication~\cite{khaskheli2022comparative}.

\textbf{Universal Design for Payment Apps:}
Most participants agreed with the implementation of a blue and white theme for the payment application. Users mentioned that they like the use of a cool and soothing color like blue for a payment application because it made them feel safer. Users also mentioned that they did not like the choice of yellow text over blue elements and expressed concerns for contrast and accessibility. Thus, universal design practices can make the app more accessible and inclusive~\cite{fogli2020universal, walker2017universal, reid2013expanding}. All users said that they found our prototype easy to navigate and liked having all the important information for a payment application presented upfront. Additionally, the use of shortcuts like a navigation bar will help make the users' navigation experience more seamless. Users mentioned that using recognized icons can help with the comprehension of the application elements. Similar to our results, prior studies have shown that a minimalist application on-boarding process increased user engagement~\cite{strahm2018generating, surendran2018emergent}.

\section{Future Work and Limitations}
\label{future_work}

As a future extension to this study, the prototype can be enhanced in two ways. First, by implementing the feedback from our participants to improve the functionality and usability of the prototype. Second, by adding extra functionality to the prototype and making it closer to a real world payment application. These improvements can facilitate a more seamless experience during user testing and overcome user bias in reporting on a prototype. Moreover, the user study can be modified to ask users to perform certain tasks on the applications. By using a think-aloud protocol and timing the users on the tasks, we can understand the pain-points of the prototype. Further, usability of the prototype can be measured using the System Usability Scale (SUS). Such improvements can lead to richer findings and a more comprehensive understanding of user perception of our prototype.

\section{Conclusion}
\label{conclusion}
In this study, we built a prototype of e-payment application using Figma and recruited $N=12$ participants for prototype testing followed by a semi-structured interview. We find that the usability of a mobile application informs important perception on the security and privacy for the users. Users pay attention to ease-of-use and aesthetics of the applications, such as use of soothing and cool colors, that make them feel safe. We provide recommendations from our study and directions for potential future extensions.

{\footnotesize \bibliographystyle{acm}
\bibliography{references}}

\appendix

\section{Pre-Screening Survey}

\textbf{Initial Screening}

\begin{enumerate}
\item	What is your age?

\item	What is your email address?

\item	Select the e-payment or mobile banking applications that you have used \/ utilize (choose at least one)  $\circ$ PayPal $\circ$ Venmo $\circ$ CashApp $\circ$ Apple Pay $\circ$ Google Pay $\circ$ Samsung Pay $\circ$ Meta Pay $\circ$ Zelle $\circ$ Banking app (please specify) $\circ$ other e-payment app (please specify)

\end{enumerate}

\textbf{Everyday Digital Usage}

\begin{enumerate}
\item	"I often ask others for help with the computer or mobile phone." 
$\circ$	Strongly Disagree  
$\circ$ Somewhat Disagree  
$\circ$ Neither Agree nor Disagree 
$\circ$ Somewhat Agree 
$\circ$ Strongly Agree

\item	"Others often ask me for help with the computer or mobile phone."  $\circ$	Strongly Disagree  $\circ$ Somewhat Disagree  $\circ$ Neither Agree nor Disagree $\circ$ Somewhat Agree $\circ$ Strongly Agree

\item	Which of the following things have you done? (multiple-select)  
$\circ$ Designed a website 
$\circ$ Designed a mobile app 
$\circ$ Registered a domain name 
$\circ$ Used SSH  
$\circ$ Configured a firewall  
$\circ$ Created a database  
$\circ$ Installed a computer program
$\circ$ Written a computer program  
$\circ$ Deployed an application on a mobile app store 
$\circ$ None of the above

\item	Do you have an up-to-date virus scanner on your computer?
$\circ$ Yes
$\circ$ No

\item	Have you taken or taught a course on computer security?
$\circ$ Yes
$\circ$ No

\item	Is computer security one of your main job responsibilities?
$\circ$ Yes
$\circ$ No

\item	Do you have a security cognate or do you plan on security being a feature of your job role?
$\circ$ Yes
$\circ$ No

\item	To your knowledge, have you ever been hacked or compromised by attackers?
$\circ$ Yes
$\circ$ No
$\circ$ I don't know

\item	Have you ever suffered data loss for any reason? (ex: Hacking, data corruption, hard drive failure)
$\circ$ Yes
$\circ$ No
$\circ$ I don't know

\item	"I can control the risk of my online account being taken over or being phased."
$\circ$	Strongly Disagree  
$\circ$ Somewhat Disagree  
$\circ$ Neither Agree nor Disagree 
$\circ$ Somewhat Agree 
$\circ$ Strongly Agree

\item	"The harm of being locked out of my account would be immediate." 
$\circ$	Strongly Disagree  
$\circ$ Somewhat Disagree  
$\circ$ Neither Agree nor Disagree 
$\circ$ Somewhat Agree 
$\circ$ Strongly Agree

\item	"If I am exposed to online risk, I can mitigate the harm." 
$\circ$	Strongly Disagree  
$\circ$ Somewhat Disagree  
$\circ$ Neither Agree nor Disagree 
$\circ$ Somewhat Agree 
$\circ$ Strongly Agree

\item	"Experts understand phishing and know how to protect me from it." 
$\circ$	Strongly Disagree  
$\circ$ Somewhat Disagree  
$\circ$ Neither Agree nor Disagree 
$\circ$ Somewhat Agree 
$\circ$ Strongly Agree

\item	"I understand phishing and know how to protect myself." 
$\circ$	Strongly Disagree  
$\circ$ Somewhat Disagree  
$\circ$ Neither Agree nor Disagree 
$\circ$ Somewhat Agree 
$\circ$ Strongly Agree

\item	"Phishing is widespread and affects many people, and not only for target individuals." 
$\circ$	Strongly Disagree  
$\circ$ Somewhat Disagree  
$\circ$ Neither Agree nor Disagree 
$\circ$ Somewhat Agree 
$\circ$ Strongly Agree

\item	"Phishing is everywhere. It is a constant exposure." 
$\circ$	Strongly Disagree  
$\circ$ Somewhat Disagree  
$\circ$ Neither Agree nor Disagree 
$\circ$ Somewhat Agree 
$\circ$ Strongly Agree

\item	"Phishing is a new kind of risk." 
$\circ$	Strongly Disagree  
$\circ$ Somewhat Disagree  
$\circ$ Neither Agree nor Disagree 
$\circ$ Somewhat Agree 
$\circ$ Strongly Agree

\end{enumerate}

\textbf{Demographics}
\begin{enumerate}

\item	What ethnicity do you identify with? (Please select all that apply)
$\circ$	American Indian or Native American   
$\circ$ Asian  
$\circ$ Native Hawaiian or Other Pacific Island
$\circ$ White  
$\circ$ Hispanic  
$\circ$ Other. Please Specify    
$\circ$ Do not Wish to Specify 

\item	Which gender do you identify with the most?
$\circ$	Female  
$\circ$ Male 
$\circ$ Transgender  
$\circ$ Non-binary  
$\circ$ A gender not listed here
$\circ$ Do not Wish to Specify

\item	What is the highest level of education you have completed? (If currently enrolled, highest degree received.)
$\circ$	Less than high school  
$\circ$	High school graduate 
$\circ$	Diploma 
$\circ$	Vocational training 
$\circ$	Bachelors degree program 
$\circ$	Masters degree program 
$\circ$	Professional degree 
$\circ$	Doctorate 
$\circ$	Other. Please Specify
\end{enumerate}

\section{Interview Questions}


\textbf{Security}

\begin{enumerate}

\item	Do you think the app is secure or does it have some vulnerabilities? What security features did you notice? What potential vulnerabilities or issues did you notice?
\item	What are your opinions on Multi-Factor Authentication (MFA) options like biometric authentication, One Time Passwords (OTP), etc. while logging in? Have you used face id for payment? What are your thoughts on that?
\item	What are your opinions on MFA options before sending money?
\item	What is your opinion of the feature that allows you to add beneficiaries using QR codes? Have you heard of this before? Have you used it before?
\item	What is your opinion on receiving login notifications? Do you take any steps when you receive them?

\end{enumerate}

\textbf{Privacy}

\begin{enumerate}

\item	Do you think the app is designed while keeping user’s privacy in mind? Why or why not?
\item	What did you think about the permissions requested by the application? Do you usually scrutinize permissions for applications, or do you just accept whatever is requested?
\item	Were you able to find Privacy Policies? Did you read them? Do you usually read and understand privacy policies?
\item	What do you think about the identity verification procedures for payment applications where you send a document like Driver’s License or SSN number to verify you as a user? Do you think that is helpful or redundant?
\item	What is your opinion on social features of e-payment the applications such as Venmo?

\end{enumerate}

\textbf{Usability}

\begin{enumerate}

\item	What is your overall opinion on the experience of using the application? Is there something you liked or disliked about it?
\item	What did you think about the theme and color scheme of the application? Is there anything you’d like to add or change?
\item	What did you think about the font used in the application? Do you like to have the ability to change font type or size?
\item	What did you think about the icons used in the application? Do you think the icons help or do you think the application would be better without them?
\item	What did you think about the Graphical User Interface (GUI) organization and layout of the application? Was it easy to use? Was it intuitive? Did it take you a long time to familiarize yourself with the application?

\end{enumerate}

\end{document}